\documentclass[10pt, jmp, eps, superscriptaddress,showpacs,superscriptaddress]{revtex4-1}
\usepackage{amsmath,amssymb,amsbsy,color,graphicx}
\newtheorem{Theorem}{\bf Theorem}
\newtheorem{Proposition}[Theorem]{\bf Proposition}
\newtheorem{Lemma}{\bf Lemma}

\newcommand{\Proof}{\vspace*{0.3em} \noindent {\bf Proof:}~}
\newcommand{\Remark}{\vspace*{1em} \noindent {\bf Remark:}~}
\newcommand{\Notation}{\vspace*{1em} \noindent {\bf Notation:}~}

\newcommand{\ave}[1]{\left\langle #1 \right\rangle}
\newcommand{\norm}[1]{\left|\left| #1 \right|\right|}
\newcommand{\dfracd}[2]{\dfrac{{\rm d}#1}{{\rm d}#2}}
\newcommand{\dfracp}[2]{\dfrac{\partial #1}{\partial #2}}

\begin{document}

\title{On the neighborhood of an inhomogeneous stable stationary solution of the Vlasov equation - Case of the Hamiltonian mean-field model}

\author{Julien Barr\'e}
\affiliation{Laboratoire J.A. Dieudonn\'e,
  Universit\'e de Nice Sophia-Antipolis,
  UMR CNRS 7351, Parc Valrose, F-06108 Nice Cedex 02, France}

\author{Yoshiyuki Y. Yamaguchi}
\affiliation{Department of Applied Mathematics and Physics,
Graduate School of Informatics, Kyoto University, Kyoto, 606-8501, Japan}

\date{\today}

\begin{abstract}
  We consider the one-dimensional Vlasov equation with an attractive
  cosine potential, and its non homogeneous stationary states that are
  decreasing functions of the energy. We show that in
  the Sobolev space $W^{1,p}$
  ($p>2$) neighborhood of such a state, all stationary states that are
  decreasing functions of the energy are stable. This is in sharp
  contrast with the situation for homogeneous stationary states of a
  Vlasov equation, where a control over strictly more than one
  derivative is needed to ensure the absence of unstable stationary
  states in a neighborhood of a reference stationary
  state~[Z.Lin and C.Zeng, Comm.Math.Phys. {\bf 306}, 291-331 (2011)].
\end{abstract}

\maketitle

\section{Introduction}

Vlasov equation is central in different areas of physics, notably plasma 
physics, where it is used with the Coulomb potential, and astrophysics, 
where the Newton potential is used instead. In this latter context, it is 
usually called ``collisionless Boltzmann equation''.
Understanding the asymptotic behavior of a perturbation to a
stationary state of the Vlasov equation is an old problem.  A huge
literature is devoted to the linearized dynamics, starting with the
pioneering work of Landau \cite{Landau46}. The full non linear problem,
despite a large literature (for
instance~\cite{BGK,ONeil65,Holloway91,Lancellotti98,Manfredi97}), is still not
fully understood.

The subject has witnessed spectacular mathematical progresses
recently.  Mouhot and Villani~\cite{Mouhot10,Mouhot11} showed that if
the initial condition is close, in some analytical norm, to a stable
homogeneous stationary state, then the dynamics is an exponential
relaxation towards another nearby stable homogeneous stationary
state. Lin and Zeng in \cite{Lin11} investigated weaker norms
of the Sobolev space $W^{s,p}$ with $p>1$.
They showed among other results that if
the norm is weak enough (precisely, $s<1+1/p$), any neighborhood of a
stable homogeneous stationary state also contains unstable homogeneous
stationary states, as well as small BGK waves. In particular, complete
damping for any initial condition, as in Mouhot-Villani's setting, is
excluded. Conversely, if $s>1+1/p$, there is a neighborhood of the
reference stable state that contains no unstable stationary
states. All these impressive results hold for homogeneous stationary
states: this is unfortunately a severe limitation, since it excludes
all situations of interest for self-gravitating systems.

A natural question is then: what could it be possible to show in the
context of non-homogeneous stationary states?  First, any exponential
relaxation as in \cite{Mouhot10,Mouhot11} is impossible, since one
always expects an algebraic relaxation, already for the
linearized problem \cite{BOY11,BY13}. One may then conjecture an
algebraic relaxation at the non linear level (see \cite{Morita11}
  for such a conjecture in the context of the 2D Euler equation), but
it seems difficult to prove. Now, is an analysis in the spirit of
\cite{Lin11} possible?  Again, the complexity of the linearized
problem is a serious obstacle (see for instance \cite{BinneyTremaine}
for a textbook account of the study of the linearized Vlasov equation
in astrophysics). However, a simple criterion for the stability of a
large class of non-homogeneous stationary states has been found
recently~\cite{Ogawa13}, in the context of a simple toy model, called
the Hamiltonian Mean-Field model (HMF).

The purpose of this paper is to take advantage of this simple
formulation to investigate the neighborhood of inhomogeneous stable
stationary solutions in the case of the HMF model. This is a first
partial advance, in the spirit of Lin and Zeng, for inhomogeneous
stationary states. We will show that the results differ significantly
from the homogeneous case: it is actually easier to rule out the
presence of unstable states in a neighborhood of the reference stable
state, since a $W^{1,p}$ norm may be sufficient.

We state precisely our results in section~\ref{sec:results},
emphasizing the important difference with the homogeneous case, and
postpone the proofs to section~\ref{sec:proofs}. Section~\ref{sec:num}
presents some numerical illustration of our findings.

\section{Statement of the results}
\label{sec:results}

The Vlasov equation associated to the HMF model is
\begin{equation}
\label{eq:vlasov}
    \dfracp{f}{t} + \{ h, f \} = 0,
    \quad {\rm with}\quad
    \{ f, g\} = \dfracp{f}{p} \dfracp{g}{q} - \dfracp{f}{q} \dfracp{g}{p}.
\end{equation}
and the one-body Hamiltonian of the HMF model is
\begin{equation}
    \label{eq:hamiltonian}
    h(q,p,t) = \dfrac{p^{2}}{2} - M[f](t)\cos (q-\varphi(t)),
    \quad
    M[f](t) e^{i\varphi(t)}= \iint_{\mu} f(q,p,t)(\cos q+i\sin q) dqdp,
\end{equation}
where $\mu$ represents the phase space of the one-body system.

\Remark
Note that $0\leq M[f]\leq 1$. Furthermore, thanks
to the rotational symmetry of the HMF model,
we may set the magnetization's phase to zero without loss of generality.
We will always do so in the following.

\Notation If $f$ is stationary, the Hamiltonian system $h$ is
integrable, and we can introduce the angle-action
variables $(\Theta,J)$.  An integrable Hamiltonian $h(q,p)$ can be
expressed as a function of $J$ only.  We denote such a Hamiltonian as
$H(J)$. We also write $\Omega(J)=\partial_{J}H(J)$.

\Remark The phase space of Hamiltonian \eqref{eq:hamiltonian} presents
a separatrix, at energy $M$. Strictly speaking, one must then define
the angle-action variables separately in the different regions
delimited by the separatrix.  This is technical and a little bit
cumbersome, so we postpone it to section~\ref{sec:angle-action}.

\vspace*{1em}
Clearly, any function $f$ that depends on $(q,p)$ through the
Hamiltonian $h$ only is a stationary solution to \eqref{eq:vlasov}. In
this paper, we concentrate on the following special class of
stationary solutions:

\begin{Definition}
  A function $f$ is called a monotonous stationary solution
  if it can be written as
\[
f(q,p) = F(h(q,p))
\]
with $F$ a $C^1$, real, strictly decreasing function,
and if it is normalized: $\iint_{\mu} fdqdp=1$.
\end{Definition}

\vspace*{1em} Note that a monotonous stationary solution $f$ is
non-homogeneous in space if and only if $M[f]\neq 0$.  We further note
that $M[f]=1$ is excluded, since $M[f]=1$ implies that $f$ is
concentrated on the $p$-axis and hence $f$ is not $C^{1}$.  As will be
clear in the following, these stationary solutions may be stable or
unstable. This is a difference with 3D self gravitating systems, where
stationary solutions that are strictly decreasing functions of the
energy are always stable (see \cite{Lemou12} for the most recent
results in this direction).

To measure the distance between two stationary solutions, we will use
the fractional Sobolev spaces $W^{s,a}$. In addition, we require that
to be close to each other, two stationary solutions must not differ
too much in their magnetization, which is rather natural. 
In the whole paper, we will use ``stable'' to mean ``formally
stable''. We can now state our main result.

\begin{Theorem}
    \label{theo:main}
    Let $f$ be a non-homogeneous stable monotonous stationary state, such that
    $f\in W^{1,a}$ with $a>2$.  
    Let $\tilde {f}$ be another monotonous stationary state
    such that $\tilde{f}\in W^{1,a}$. Then
    there exists $\epsilon>0$ such that:
\begin{displaymath}
    \norm{f-\tilde{f}}_{W^{1,a}}<\varepsilon
    ~~\mbox{and}~~|M[f]-M[\tilde{f}]|<\varepsilon
    \text{~~imply that ~} 
    \tilde{f} ~\mbox{is stable}.
\end{displaymath}
In other words, there exists a neighborhood of $f$
in the $W^{1,a}(a>2)$ norm that does not contain any 
unstable monotonous stationary state, with magnetization close to $M[f]$.
\end{Theorem}
This is to be contrasted with the following statement concerning
homogeneous stationary states:
\begin{Theorem}
\label{theo:hom}
Let $f$ be a homogeneous stable monotonous stationary state, such that
$f\in W^{s,a}$, with $a>1$ and $s<1+1/a$.  Any neighborhood of $f$ in
the $W^{s,a}$ norm contains an unstable monotonous homogeneous
stationary state.
\end{Theorem}

From Theorem~\ref{theo:main}, we see that using a norm that controls
only one derivative of the distribution function is enough to ensure
that a neighborhood of $f$ is ``simple'', in the sense that it does
not contain any unstable monotonous stationary state. By contrast, in
the homogeneous case, even requiring more regularity (with $s>1$) may
not be enough.

We have stated Theorem~\ref{theo:hom} in this way to emphasize the
contrast with the non-homogeneous case. It is actually a much weaker
and less general statement of the results in~\cite{Lin11}. We will
give a proof of it, because it is instructive and for self-consistency
of the paper.

\vspace*{1em}
\noindent
{\bf Idea of the proof of Theorem~\ref{theo:main}:}
The proof relies on the analysis of the simple formal stability
criterion obtained for non-homogeneous monotonous stationary states
in~\cite{Ogawa13}.
From the condition $|M[f]-M[\tilde{f}]|<\epsilon$,
we may choose a small enough $\epsilon$ such that
$M[\tilde{f}]$ is not zero,
and hence we will assume that $\tilde{f}$ is non-homogeneous in the proof.

\Notation We need to define the average over the angle $\Theta$
variable, at fixed action $J$; for a function $A(\Theta,J)$, we denote
it as (see Sec.~\ref{sec:angle-action} for a more precise definition
of the integrals over $\Theta$ and/or $J$):
\begin{equation}
   \label{eq:theta-ave}
    \ave{A}_{J}
    = \dfrac{1}{2\pi} \int_{-\pi}^{\pi} A(\Theta,J) d\Theta.
\end{equation}

Following \cite{Ogawa13}, we now introduce the functional $I[f]$:
\begin{equation}
    \label{eq:stability-1}
    I[f] = 1 + \int \cos^{2}q dq \int \dfrac{1}{p} \dfracp{f}{p} dp
    - 2\pi \int \dfrac{1}{\Omega(J)} \dfracd{(F\circ H)}{J}
   \ave{\cos Q}_J^2 dJ,
\end{equation}
where $Q(\Theta,J)$ is the position variable $q$ in angle-action
coordinates. From~\cite{Ogawa13}, we have the convenient formal
stability criterion:
\begin{Proposition}
Let $f$ be a monotonous stationary solution. Then $I[f]>0$ if and only if $f$ 
is formally stable.
\end{Proposition}
If $M=0$, $f$ is homogeneous, and action/angle variables coincides
with the $(q,p)$ variables. Hence the average over $\Theta$ coincides
with the average over the spatial variable $q$, and $I$
is simplified to:
\begin{equation}
    \label{eq:stability-0}
    I[f] = 1 + \pi \int \dfrac{1}{p} \dfracp{f}{p} dp.
\end{equation}

With criterion \eqref{eq:stability-1} in hand, we only have to show that
if $f$ is a non-homogeneous monotonous stationary state such that $I[f]>0$, and
$\tilde{f}$ and $f$ are close in $W^{1,a}$ norm and their magnetization are 
close, then $|I[\tilde{f}]-I[f]|$ is small. 

To prove Theorem~\ref{theo:hom}, it is enough to construct, for each
$f$ homogeneous monotonous stable stationary state, a nearby (in
$W^{s,a}$ norm) homogeneous monotonous stationary state  $\tilde{f}$, 
such that $I[\tilde{f}]<0$.

\vspace{1em}
\noindent
{\bf Discussion:} 
Clearly, the proof of Theorem~\ref{theo:main} heavily relies on the
stability criterion obtained in~\cite{Ogawa13}. Thus, it is not clear
how to generalize this result to more general models than HMF.
Indeed, in general the linear stability analysis of a non-homogeneous
stationary state is complicated; see for instance~\cite{BinneyTremaine} for the
three dimensional gravitational case, but note that the situation is
not much better for general one dimensional systems.

\section{Proofs}
\label{sec:proofs}

\subsection{Proof of theorem~\ref{theo:hom}}
We first give for consistency a proof of Theorem~\ref{theo:hom},
although it is contained (in a much more general form)
in~\cite{Lin11}.  The proof relies on the stability
functional~\eqref{eq:stability-0}: given a stable monotonous
stationary state $f_0$ (hence $I[f_0]>0$), we have to find a
modification $f_1$, small in $W^{s,a}$ norm, such that $I[f_0+f_1]<0$.
Following the strategy of~\cite{Lin11}, we introduce
$g(p)=e^{-p^2/2}/(2\pi)^{3/2}$, and
$g_{\varepsilon,\alpha}(p)=\varepsilon g(p/\varepsilon^\alpha)$.
Note that
$\int g_{\varepsilon,\alpha}dqdp=\varepsilon^{1+\alpha}$. 
It is easy to see that 
\begin{equation}
    \label{eq:funcg}
    \int \frac{g^{\prime}_{\varepsilon,\alpha}(p)}{p}dp
    = - \dfrac{1}{2\pi} \varepsilon^{1-\alpha} .
\end{equation}
Furthermore, for small $\varepsilon$ and $1-\alpha+\alpha/a>0$, we
have the estimate~\cite{Lin11}
\begin{equation}
    \label{eq:normsa}
    ||g_{\varepsilon,\alpha}||_{W^{s,a}} =O(\varepsilon^{1-s\alpha+\alpha/a}) .
\end{equation}
We now choose a modified state as
\begin{equation}
    f_0(p)+f_1(p)
    =\frac{1}{1+\varepsilon^{1+\alpha}}\left(f_0(p)+g_{\varepsilon,\alpha}(p) 
    \right) ,
\end{equation}
which corresponds to a modification
\begin{equation}
    \label{eq:f1}
    f_1(p)
    = \frac{1}{1+\varepsilon^{1+\alpha}}g_{\varepsilon,\alpha}(p)-
    \frac{\varepsilon^{1+\alpha}}{1+\varepsilon^{1+\alpha}}f_0(p) .
\end{equation}
From~\eqref{eq:stability-0} and~\eqref{eq:funcg}, it is clear that 
$g_{\varepsilon,\alpha}$ induces a large negative variation of the 
stability functional as soon as $\alpha>1$. Hence in this case 
$I[f_0+f_1]<0$, and $f_0+f_1$ is unstable.

From the expression of $f_1$~\eqref{eq:f1}, we see that the only way
$||f_1||_{W^{s,a}}$ could be large is if
$||g_{\varepsilon,\alpha}||_{W^{s,a}}$ itself is large.
Now, from~\eqref{eq:normsa}, $g_{\varepsilon,\alpha}$ is small in $W^{s,a}$
norm if $1-s\alpha+\alpha/a>0$.
We see that it is possible to choose
$\alpha$ such that the two conditions $\alpha>1$ and
$1-s\alpha+\alpha/a>0$ are satisfied, as soon as $s<1+1/a$.
Remembering that \eqref{eq:normsa} is valid
for $1-\alpha+\alpha/a>0$, $\alpha>1$ implies $a>1$.

This completes the proof of Theorem~\ref{theo:hom}. $\blacksquare$

\subsection{Angle-action variables}
\label{sec:angle-action}

We need to define a bijection between position/momentum $(q,p)$ and
angle/action $(\Theta,J)$ coordinates. We will repeatedly use this
change of variable in both directions. To keep notations as understandable as possible, we will use the
following convention: functions of $(q,p)$ will be denoted with small
letters (for instance $q,p,h(q,p),j(q,p)\ldots$), and functions of
$(\Theta,J)$ with capital letters (for instance $\Theta, J,
H(J),Q(\Theta,J)\ldots$).

\begin{figure}
    \includegraphics[width=8cm]{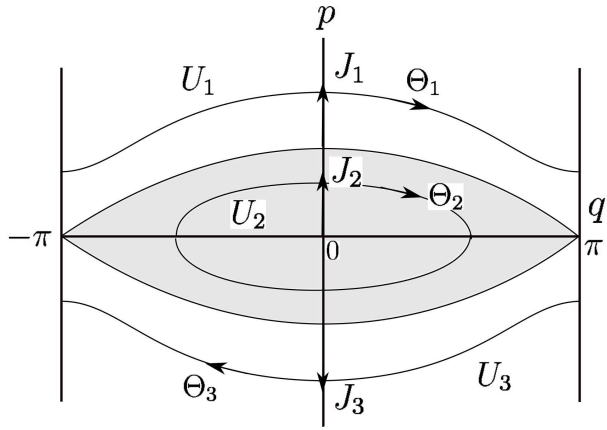}
    \caption{The one particle phase space $\mu$, divided in
      the three regions $U_1,U_2$ and $U_3$.}
    \label{fig:angle-action}
\end{figure}

As a further difficulty, the presence of a separatrix in phase space
imposes us to divide the phase space in three regions, in order to properly 
define the change of variables: above separatrix ($U_1$),
inside the separatrix ($U_2$) and below separatrix ($U_3$), see 
Fig.~\ref{fig:angle-action}. 
In equations:
\begin{eqnarray}
    U_1&=& \{(q,p)~|~ h(q,p)>M,~p>0\}, \nonumber \\
    U_2&=& \{(q,p)~|~ -M<h(q,p)<M \}, \nonumber \\
    U_3&=& \{(q,p)~|~ h(q,p)>M,~p<0\}. \nonumber
\end{eqnarray}
Over each of these three regions, it is possible to define a bijective
change of variables 
\begin{eqnarray}
    U_i &\to& V_i \nonumber \\
    (q,p) &\mapsto& (\theta_i(q,p),j_i(q,p)) \nonumber
\end{eqnarray}
with 
\begin{eqnarray}
    V_1&=& \{(\Theta_{1},J_{1})~|~\Theta_{1}\in ]-\pi,\pi],~J_{1}>4\sqrt{M}/\pi \}, \nonumber \\
    V_2&=& \{(\Theta_{2},J_{2})~|~\Theta_{2}\in ]-\pi,\pi],~0<J_{2}< 8\sqrt{M}/\pi \},
    \nonumber \\
    V_3&=& \{(\Theta_{3},J_{3})~|~\Theta_{3}\in ]-\pi,\pi],~J_{3} >4\sqrt{M}/\pi \}.
\end{eqnarray}
The inverse change of variables reads
\begin{eqnarray}
    V_i &\to& U_i \nonumber \\
    (\Theta_{i},J_{i}) &\mapsto& (Q_i(\Theta_{i},J_{i}),P_i(\Theta_{i},J_{i})).\nonumber
\end{eqnarray}
To keep notations simple, we will however use a single
notation for each of these functions:
$\theta(q,p), j(q,p), Q(\Theta,J), P(\theta,J)$.  Similarly, any real
function $G$ of the angle-action variables is thus actually made of
three distinct functions
\[
G_i: V_i \to \mathbb{R}\quad i=1,2,3. 
\]
We will however use for such a function a single notation $G(\Theta,J)$.
The integrals over $d\Theta~dJ$ are thus to be understood as the sum 
of three integrals over $V_1,V_2$ and $V_3$:
\[
\iint_{\mu} G(\Theta,J) d\Theta dJ
= \sum_{i=1}^{3} \iint_{V_{i}} G_{i}(\Theta_{i},J_{i}) d\Theta_{i} dJ_{i}.
\]
The average over $\Theta$ defined in \eqref{eq:theta-ave} also yields 
three functions of the action, which we do not write explicitly. 

\subsection{General strategy}
\vspace*{1em} For later use, we rewrite the stability functional
\eqref{eq:stability-1} to make it easier to analyze.

\begin{Lemma}
    \label{lem:I-rewrite}
    Let $f$ be a monotonous stationary solution.
    The stability functional $I[f]$ \eqref{eq:stability-1}
    can be rewritten as
    \begin{equation}
        \label{eq:stability-2}
        \begin{split}
            I[f]
            = 1 + \iint_{\mu} F'(h(q,p))  w(q,p) dqdp,
        \end{split}
    \end{equation}
 with
 \begin{equation}
     \label{eq:w}
     w(q,p) = \ave{\cos^2 Q}_{j(q,p)}-\ave{\cos Q}_{j(q,p)}^{2} .
 \end{equation}
 Note that the function $w$ implicitly depends on $f$ through the
 definitions of the functions $Q$ and $j$.
\end{Lemma}

\Proof
Remembering $F'(h(q,p))=F'(H(J))$,
the second term of \eqref{eq:stability-1} is:
\begin{eqnarray}
    \label{eq:Fdashcos2}
    &&\iint_{\mu} F'(h(q,p)) \cos ^2q dqdp
    = \int dJ~ F'(H(J)) \int \cos^2 Q(\Theta,J)d\Theta \nonumber \\
    &=& \iint_{\mu} F'(H(J)) \ave{\cos^2 Q}_J d\Theta dJ
    = \iint_{\mu} F'(h(q,p)) \ave{\cos^2 Q}_{j(q,p)} dqdp.
\end{eqnarray}
Similarly, the third term is
\begin{equation}
    \begin{split}
        & - 2\pi \int F'(H(J)) \ave{\cos Q}_{J}^{2}dJ
        = -\iint_{\mu} F'(H(J)) \ave{\cos Q}_{J}^{2} d\Theta dJ
        = - \iint_{\mu} F'(h(q,p)) \ave{\cos Q}_{j(q,p)}^{2} dqdp.
        \quad \blacksquare
    \end{split}
\end{equation}

\Remark Looking back at \eqref{eq:Fdashcos2} and using the fact
$Q(\theta(q,p),j(q,p))=q$, we may replace the function $w(q,p)$
defined in \eqref{eq:w} with
\begin{equation}
    w_{1}(q,p) = \cos^{2}q - \ave{\cos Q}^{2}_{j(q,p)}.
\end{equation}

We consider $f=F\circ h$ a stable non-homogeneous monotonous 
stationary state, and $\tilde{f}=\tilde{F}\circ\tilde{h}$
another monotonous stationary state.
$\tilde{h}$ is the Hamiltonian corresponding to $\tilde{f}$:
\begin{equation}
    \label{eq:tilde-hamiltonian}
    \tilde{h}(q,p) = \dfrac{p^{2}}{2} - \tilde{M}\cos q,
    \quad
    \tilde{M} = \iint_{\mu} \tilde{f}(q,p)\cos q dqdp,
\end{equation}
and the angle-action variables associated to $\tilde{h}$ are written
$(\tilde{\Theta},\tilde{J})$.
The change of variable is $(\tilde{\theta}(q,p),\tilde{j}(q,p))$,
and the inverse change is 
$(\tilde{Q}(\tilde{\Theta},\tilde{J}),\tilde{P}(\tilde{\Theta},\tilde{J}))$.
The stability functional for $\tilde{f}$ is 
\begin{equation}
    \label{eq:stability-2-tilde}
    I[\tilde{f}]
    = 1 + \iint_{\mu} \tilde{F}'(\tilde{h}(q,p))\left[ \ave{\cos^2 \tilde{Q}}_{\tilde{j}(q,p)}-\ave{\cos \tilde{Q}}_{\tilde{j}(q,p)}^2\right] dqdp.
\end{equation}
We write 
\[
\tilde{w}(q,p) = \ave{\cos^2 \tilde{Q}}_{\tilde{j}(q,p)}-\ave{\cos \tilde{Q}}_{\tilde{j}(q,p)}^2.
\]
Since $f$ is stable, $I[f]>0$. Thus, to prove 
Theorem~\ref{theo:main}, it is enough to show that if\\ 
{\bf H1.} $||\tilde{f}-f||_{W^{1,a}}$ is small; and\\ 
{\bf H2.} $|\tilde{M}-M|$ is small, then
\begin{equation}
    \left| I[\tilde{f}]-I[f] \right| ~{\rm is~small.} 
\end{equation}

\vspace*{1em}
For convenience we denote the discrepancies by
\begin{equation}
    \Delta M = \tilde{M}-M \quad;\quad \Delta I = I[\tilde{f}] - I[f].
\end{equation}
$\Delta I$ can be rewritten as
\begin{equation}
    \Delta I
    = \iint_{\mu} \left[ (\tilde{F}'\circ\tilde{h})\tilde{w}
      - (F'\circ h) w \right] dqdp
    = \Delta I_{1} - \Delta I_{2}
\end{equation}
where
\begin{equation}
    \label{eq:DeltaI1}
    \Delta I_{1}
    = \iint_{\mu} \left[
          \tilde{F}'\circ\tilde{h} - F'\circ h \right]
        \tilde{w} dqdp
\end{equation}
and
\begin{equation}
    \label{eq:DeltaI2}
    \Delta I_{2}
    = \iint_{\mu} (F'\circ h) \left[
      \ave{\cos \tilde{Q}}_{\tilde{j}(q,p)}^{2} - \ave{\cos Q}_{j(q,p)}^{2}
    \right] dqdp.
\end{equation}
We have used here the remark after Lemma \ref{lem:I-rewrite}.
We have
\begin{equation}
    |\Delta I| \leq |\Delta I_{1}| + |\Delta I_{2}|,
\end{equation}
and will show smallness of $|\Delta I_{1}|$ and $|\Delta I_{2}|$
in Secs.~\ref{sec:DeltaI1} and \ref{sec:DeltaI2} respectively.

\subsection{$|\Delta I_{1}|$ is small}
\label{sec:DeltaI1}
In this section, the hypothesis {\bf H1} on
$||\tilde{f}-f||_{W^{1,a}}$ will be crucial;
we will also use {\bf H2}.
We begin with some Lemmas.
\begin{Lemma}
    \label{lem:Qa-converge}
    Let $m$ be a positive constant
    and the function $u_{a}$ be defined by
    \begin{equation}
        u_{a}(q,p;m) = \left( |p|^{a} + |m\sin q|^{a} \right)^{1/a}.
    \end{equation}
    Then, $\norm{1/u_{a}}_{L^{b}}$ is finite
    for any $a>0$ and $1<b<2$.
 Moreover, when $m$ is small, the leading order is $O(m^{-1/b})$.
\end{Lemma}

\Proof
The considered norm is
\begin{equation}
    \norm{\dfrac{1}{u_{a}} }_{L^{b}}^{b}
    = \iint_{\mu} \dfrac{dqdp}{(|p|^{a}+|m\sin q|^{a})^{b/a}}.
\end{equation}
We have to check the convergence of the integral at the points
where the integrand diverges, which are $(q,p)=(0,0),~(\pm\pi,0)$,
and when $|p|\to\infty$. 
\begin{itemize}
      \item Around $(0,0)$ and $(\pi,0)$:\\
    Let $(q_{0},0)$ be the point we are considering.
    Using polar coordinates:
    $q-q_{0}=\dfrac{r}{m}\cos\rho,~p=r\sin\rho$,
    we have $dqdp=\dfrac{r}{m}drd\rho$ and
    \begin{equation}
    \dfrac{dqdp}{(|p|^{a}+|m\sin q|^{a})^{b/a}} = \dfrac{1}{m} \dfrac{d\rho}{(|\sin\rho|^{a}+|\cos\rho|^{a})^{b/a}}r^{1-b} dr
    \end{equation}    
    The integral over $\rho$ is finite, and the integral over $r$
    converges for $b<2$.
      \item $|p|\to\infty$: 
\begin{equation}     
 \dfrac{dqdp}{(|p|^{a}+|m\sin q|^{a})^{b/a}} \leq \dfrac{dp}{|p|^{b}}
 \end{equation}    
 Thus the integral converges  for $1<b$.
\end{itemize}
Putting all together, we conclude that the integral converges
for $1<b<2$.
Moreover, if $m$ is small, the leading order of $\norm{1/u_{a}}_{L^{b}}$
is $O(1/m^{1/b})$ from the estimations around $(0,0)$ and $(\pi,0)$.
$\blacksquare$

\Remark
That $m$ is non-zero is important to ensure convergence for $1<b<2$.
If $m=0$, then the integrand diverges around $(0,0)$ only,
but the divergence occurs on the line $p=0$. As a result,
around the line $p=0$,
\begin{equation}
    \iint \dfrac{dqdp}{(|p|^{a})^{b/a}}
    \simeq \int \dfrac{dp}{|p|^{b}}
\end{equation}
which converges for $b<1$. Considering the estimation for
$|p|\to\infty$, which requires $1<b$, the interval of $b$ for
convergence is empty!

Since $|\cos Q|\leq 1$, $|\tilde{w}| \leq 1$. Hence
\begin{equation}
    |\Delta I_{1}|
    \leq \iint_{\mu} \left| \tilde{F}'\circ\tilde{h}-F'\circ h \right| dqdp
    = \norm{\tilde{F}'\circ\tilde{h}-F'\circ h}_{L^{1}}.
\end{equation}
Since $M>0$, we introduce $u_{a}(q,p;M)$,
and the H\"older inequality leads to:
\begin{equation}
    |\Delta I_{1}|
    \leq \norm{(\tilde{F}'\circ \tilde{h} - F'\circ h) u_{a} }_{L^{a}}
    \norm{\dfrac{1}{u_{a}}}_{L^{b}}
\end{equation}
with $a$ and $b$ non-negative real numbers such that $1/a+1/b=1$.
The norm $\norm{1/u_{a}}_{L^{b}}$ is finite for $1<b<2$
by Lemma \ref{lem:Qa-converge},
and our job is to show that
\begin{equation}
    \norm{(\tilde{F}'\circ \tilde{h} - F'\circ h) u_{a} }_{L^{a}}^{a}
    = \iint_{\mu} \left| \tilde{F}'\circ \tilde{h} - F'\circ h \right|^{a} \left( |p|^{a} + |M\sin q|^{a} \right) dqdp
\end{equation}
is small. The first term is rewritten as
\begin{equation}
    \iint_{\mu} \left| \tilde{F}'\circ \tilde{h} - F'\circ h \right|^{a} |p|^{a} dqdp
    = \iint_{\mu} \left| \partial_{p}\left( \tilde{f} - f \right) \right|^{a} dqdp
    \leq \norm{\tilde{f}-f}_{W^{1,a}},
\end{equation}
and is small by the hypothesis $\norm{\tilde{f}-f}_{W^{1,a}}$ small.
Using the trick:
\begin{equation}
    M\sin q \left( \tilde{F}'\circ\tilde{h} - F'\circ h \right)
    = \partial_{q} \left( \tilde{f} - f \right)
    - \Delta M \sin q \tilde{F}'\circ\tilde{h},
\end{equation}
we rewrite the second term as
\begin{equation}
    \begin{split}
        & \iint_{\mu} \left| \tilde{F}'\circ\tilde{h} - F'\circ h \right|^{a} |M\sin q|^{a} dqdp
        = \iint_{\mu} \left| \partial_{q} \left( \tilde{f} - f \right)
          - \Delta M \sin q \tilde{F}'\circ\tilde{h}  \right|^{a} dqdp \\
        & \leq 2^{a} \max\left[
          \iint \left| \partial_{q} \left( \tilde{f} - f \right) \right| dqdp,
          ~
          \left|\Delta M\right|^{a}
          \iint \left| \sin q \tilde{F}'\circ\tilde{h}  \right|^{a} dqdp
        \right] \\
        & \leq 2^{a} \max\left[ \norm{\tilde{f}-f}_{W^{1,a}},~
          \dfrac{\left|\Delta M\right|^{a}}{M^{a}}
          \iint \left| \partial_{q} \tilde{f} \right|^{a} dqdp
        \right].
    \end{split}
\end{equation}
Thus, the second term is also small
by the hypothesis $\norm{\tilde{f}-f}_{W^{1,a}}$ small,
$|\Delta M|$ small and $\tilde{f}\in W^{1,a}$.
We have therefore proven that $|\Delta I_{1}|$ is small
using the main hypotheses {\bf H1} and {\bf H2}.
$\blacksquare$

\subsection{$|\Delta I_{2}|$ is small}
\label{sec:DeltaI2}

In this section, the crucial hypothesis is {\bf H2},
i.e. $|\Delta M|$ is small.
If $\Delta M=0$, then
\begin{equation}
    \begin{split}
        \tilde{M}=M
        \quad\Longrightarrow\quad
        \tilde{h}=h
        \quad\Longrightarrow\quad
        (\tilde{\theta},\tilde{j}) = (\theta,j)
        \text{~and~} (\tilde{Q},\tilde{P})=(Q,P)
        \quad\Longrightarrow\quad
        \Delta I_{2} = 0,
    \end{split}
\end{equation}
so that $\Delta I_2$ is trivially small.
We therefore consider the case $\Delta M\neq 0$.
We may choose $\Delta M>0$ without loss of generality, 
since we can exchange the roles of $f$ and $\tilde{f}$
in order to estimate $|\Delta I_{2}|$ when $\Delta M<0$.

The quantity $\Delta I_{2}$, which reads
\begin{equation}
    \Delta I_{2} = \iint_{\mu} (F'\circ h)
      \left[ \ave{\cos \tilde{Q}}_{\tilde{j}(q,p)}^{2} - \ave{\cos Q}_{j(q,p)}^{2}
      \right] dqdp,
\end{equation}
depends on $\tilde{f}$ only through the Hamiltonian and magnetization,
while $\Delta I_{1}$ directly depends on the derivative of
$\tilde{F}$.  Thus, it is rather natural to expect that $\Delta I_{2}$
is small if $\tilde{M}$ is close to $M$.  A technical problem is that
the separatrix changes as the magnetization changes, so that a direct
comparison between $\ave{\cos \tilde{Q}}_{\tilde{j}(q,p)}$ and
$\ave{\cos Q}_{j(q,p)}$ becomes difficult around the separatrix. To
solve this problem, we divide the $\mu$ space into three regions:
\begin{enumerate}
      \item Inside the separatrix
      \item Close to the separatrix
      \item Outside the separatrix
\end{enumerate}
For this purpose, we introduce $M_{1}$ and $M_{2}$ as
\begin{equation}
    \label{eq:M1M2}
     M_{1}=M-2\Delta M, \qquad
     M_{2}=M+2\Delta M.
\end{equation}
From {\bf H2}, we may consider a small $\Delta M$ which makes
$M_{1}$ positive. Thus, we also have $\tilde{M}>0$.

We now use the following strategy.  $\Delta I_{2}$ is divided in three
parts, according to the division of the $\mu$ space detailed below. 
Secs.\ref{sec:mu2},\ref{sec:mu1} and
\ref{sec:mu3} show the smallness of the contribution to $\Delta I_2$
of the region close to, inside, and outside the separatrix
respectively.

\subsubsection{Division of $\mu$ space}
\label{sec:division}

Using the Hamiltonian
\begin{equation}
    h(q,p) = \dfrac{p^{2}}{2} - M\cos q,
\end{equation}
and $M_{1},M_{2}$ defined in \eqref{eq:M1M2}, we divide $\mu$
into three $\mu_{j}$ as
\begin{equation}
    \mu = \mu_{1} \cup \mu_{2} \cup \mu_{3}
\end{equation}
where
\begin{equation}
    \begin{split}
        & \mu_{1} = \{ (q,p)\in\mu~|~ h(q,p) < 2M_{1}-M \} \\
        & \mu_{2} = \{ (q,p)\in\mu~|~ 2M_{1}-M\leq h(q,p) \leq 2M_{2}-M \} \\
        & \mu_{3} = \{ (q,p)\in\mu~|~ 2M_{2}-M < h(q,p) \}. \\
    \end{split}
\end{equation}
Accordingly, the second term $\Delta I_{2}$ is divided as
\begin{equation}
    \Delta I_{2} = \Delta I_{21} + \Delta I_{22} + \Delta I_{23},
\end{equation}
where $\Delta I_{2j}$ corresponds to the integral over $\mu_{j}$:
\begin{equation}
    \Delta I_{2j} = \iint_{\mu_{j}}
    (F'\circ h)
    \left[ \ave{\cos \tilde{Q}}_{\tilde{j}(q,p)}^{2} - \ave{\cos Q}_{j(q,p)}^{2}
    \right] dqdp.
\end{equation}

\begin{figure}[h]
    \centering
    \vspace*{0.0em}
    \includegraphics[width=10cm]{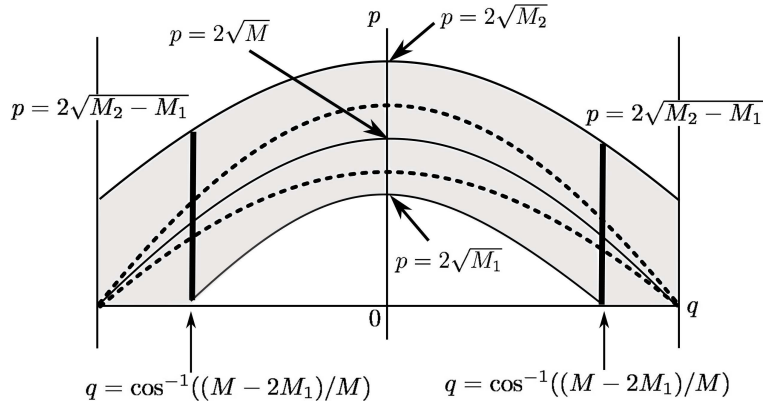}
    \caption{
        Division of $\mu$ space.
      This figure describes the upper half $\mu$ space
      and the lower half $\mu$ is similarly divided thanks to the symmetry
       $p\to -p$.
      The shaded area is $\mu_{2}$.
      The solid curves represent
      $h(q,p)=2M_{2}-M,M$ and $2M_{1}-M$ from top to bottom,
      and the dashed curves are separatrices
      for $M+|\Delta M|$ and $M-|\Delta M|$ from top to bottom.
      The vertical solid lines at $q=\cos^{-1}((M-2M_{1})/M)$
      are used to further divide $\mu_{2}$.}
    \label{fig:muspace}
\end{figure}

\subsubsection{Near the separatrix: $\mu_{2}$}
\label{sec:mu2}

\begin{Proposition}
\label{prop:DeltaI22}
Under the hypotheses of Theorem~\ref{theo:main}, $|\Delta I_{22}|$ is small.
\end{Proposition}

We first show that the area of $\mu_{2}$ is small.
\begin{Lemma}
    \label{lem:mu2-to-0}
    Let $A=\iint_{\mu_{2}} dqdp$.
Then, $A$ is estimated as $A\leq 16\pi\sqrt{\Delta M}$.
\end{Lemma}

\Proof
Introducing $q_{\rm max}=\cos^{-1}((M-2M_{1})/M)$,
we further divide $\mu_{2}$ into two parts,
\begin{equation}
    \begin{split}
        & \mu_{21} = \{ (q,p) \in \mu_{2}~|~ |q|\leq q_{\rm max} \} \\
        & \mu_{22} =  \{ (q,p) \in \mu_{2}~|~ |q|> q_{\rm max} \} \\
    \end{split}
\end{equation}
and denote the areas of $\mu_{21}$ and $\mu_{22}$ as $A_{1}$ and $A_{2}$
respectively.

The upper and the lower bound of the upper half of $\mu_{21}$
is expressed as
\begin{equation}
    p_{\rm u} = \sqrt{2(2M_{2}-M+M\cos q)},
    \quad
    p_{\rm l} = \sqrt{2(2M_{1}-M+M\cos q)}
\end{equation}
for $|q|<q_{\rm max}$.
The height of $\mu_{21}$ for a fixed $q$ is estimated as
\begin{equation}
    \begin{split}
        p_{\rm u}-p_{\rm l}
        & = \dfrac{p_{\rm u}^{2}-p_{\rm l}^{2}}{p_{\rm u}+p_{\rm l}}
        = \dfrac{4(M_{2}-M_{1})}
        {\sqrt{2(2M_{2}-M+M\cos q)}+\sqrt{2(2M_{1}-M+M\cos q)}} \\
        & \leq \dfrac{4(M_{2}-M_{1})}
        {\sqrt{2(2M_{2}-M+M-2M_{1})}+0}
        = 2\sqrt{M_{2}-M_{1}}
    \end{split}
\end{equation}
and hence the area $A_{1}$ is
\begin{equation}
    A_{1} = |\mu_{21}| \leq 2\cdot 2 \sqrt{M_{2}-M_{1}}~ 2q_{\rm max}
    = 8\sqrt{M_{2}-M_{1}}~ q_{\rm max}
\end{equation}
In $\mu_{22}$, one peace of the region is smaller than the rectangle
whose vertices are $(q_{\rm max},0),(\pi,0),(\pi,2\sqrt{M_{2}-M_{1}})$
and $(q_{\rm max},2\sqrt{M_{2}-M_{1}})$. Thus, $A_{2}$ is bounded as
\begin{equation}
    A_{2} = |\mu_{22}| \leq 4\cdot 2\sqrt{M_{2}-M_{1}} (\pi-q_{\rm max})
    = 8\sqrt{M_{2}-M_{1}} (\pi-q_{\rm max}).
\end{equation}
The total area $A$ is therefore
\begin{equation}
    A \leq 8\pi \sqrt{M_{2}-M_{1}}
    \leq 16\pi \sqrt{\Delta M}.
    \quad \blacksquare
\end{equation}

\noindent
{\bf Proof of Proposition \ref{prop:DeltaI22} : }
From the fact $|\cos q|\leq 1$, we have $\left|\ave{\cos
    \tilde{Q}}_{\tilde{j}}^{2}-\ave{\cos Q}_{j}^{2}\right|\leq 1$.
$F'$ is continuous, so that it is bounded by $F'_{\rm max}<+\infty$ in
a neighborhood of the separatrix, containing $\mu_2$ for $\Delta M$
small. Thus, we have
\begin{equation}
    |\Delta I_{22}|
    \leq \iint_{\mu_{2}} |F'(h(q,p))|~
    \left|
      \ave{\cos \tilde{Q}}_{\tilde{j}(q,p)}^{2} - \ave{\cos Q}_{j(q,p)}^{2}
    \right| dqdp
    \leq F'_{\rm max} A.
\end{equation}
Using Lemma \ref{lem:mu2-to-0} and {\bf H2}, we conclude that
$|\Delta I_{22}|$ is small.

\subsubsection{Inside the separatrix: $\mu_{1}$}
\label{sec:mu1}

\begin{Proposition}
\label{prop:DeltaI21}
Under the hypotheses of Theorem~\ref{theo:main}, $|\Delta I_{21}|$ is small.
\end{Proposition}

\Proof
From
$\left|\ave{\cos \tilde{Q}}_{\tilde{j}(q,p)}+\ave{\cos Q}_{j(q,p)}\right|\leq 2$,
we estimate $\Delta I_{21}$ as
\begin{equation}
    \begin{split}
        |\Delta I_{21}|
        & \leq 2 \norm{ (F'\circ h)
          \left( \phi_{\rm in}(q,p;\tilde{M})-\phi_{\rm in}(q,p;M) \right) }_{L^{1}(\mu_{1})} \\
        & \leq 2 \norm{F'\circ h}_{L^{1}(\mu_{1})}
        \sup_{(q,p)\in\mu_{1}} \left| \phi_{\rm in}(q,p;\tilde{M})-\phi_{\rm in}(q,p;M) \right|.
    \end{split}
\end{equation}
Here, we have introduced the following functions to simplify the notations:
\begin{equation}
    \ave{\cos Q}_{j(q,p)} = \phi_{\rm in}(q,p;M),
    \qquad
    \ave{\cos \tilde{Q}}_{\tilde{j}(q,p)} = \phi_{\rm in}(q,p;\tilde{M})
\end{equation}
where 
\begin{equation}
    \phi_{\rm in}(q,p;M) = \varphi_{\rm in}(\psi(q,p;M)),
\end{equation}
\begin{equation}
    \label{eq:varphi-in}
    \varphi_{\rm in}(k) = \dfrac{2E(k)}{K(k)} - 1,
\end{equation}
and
\begin{equation}
    k = \psi(q,p;M) = \sqrt{\dfrac{p^{2}/2+M(1-\cos q)}{2M}}.
\end{equation}
The functions $K(k)$ and $E(k)$ are the complete elliptic integrals
of the 1st and the 2nd kinds respectively.

The proof is done by the following three steps:
\begin{enumerate}
      \item We show that $\norm{F'\circ h}_{L^{1}(\mu_{1})}$ is finite.
    [Lemma \ref{lem:W1a-to-L1}]
      \item We extract the small $\Delta M$
    from $\phi_{\rm in}(q,p;\tilde{M})-\phi_{\rm in}(q,p;M)$.
    [Lemma \ref{lem:Taylor}]
      \item We show that the remaining supremum part is finite.
    [Lemma \ref{lem:sup-I21}]
\end{enumerate}
The three following Lemmas prove the Proposition \ref{prop:DeltaI21}.

\begin{Lemma}
    \label{lem:W1a-to-L1}
    \hspace*{1em} $F'\circ h \in L^{1}$ (this implies of course that
    $F'\circ h \in L^{1}(\mu_1)$).
\end{Lemma}

\Proof
Using the function $u_{a}(q,p;M)$ and the H\"older inequality, we have
\begin{equation}
    \norm{F'\circ h}_{L^{1}}
    \leq \norm{(F'\circ h)u_{a}}_{L^{a}} \norm{\dfrac{1}{u_{a}}}_{L^{b}},
\end{equation}
where $1\leq a,b\leq \infty$ and $1/a+1/b=1$.
By Lemma \ref{lem:Qa-converge}
the factor $\norm{1/u_{a}}_{L^{b}}$ converges for $1<b<2$,
which corresponds to $2<a<\infty$. On the other hand, we have
\begin{equation}
    \begin{split}
        \norm{(F'\circ h)u_{a}}_{L^{a}}
        & = \left( \iint |F'\circ h|^{a} (|p|^{a}|+|M\sin q|^{a}) dqdp \right)^{1/a} \\
        & = \left( \norm{\partial_{p}f}_{L^{a}}^{a} + \norm{\partial_{q}f}_{L^{a}}^{a} \right)^{1/a}
        \leq \norm{f}_{W^{1,a}}.
    \end{split}
\end{equation}
Thus, we have
\begin{equation}
    \norm{F'\circ h}_{L^{1}} \leq 
    \norm{f}_{W^{1,a}} \norm{\dfrac{1}{u_{a}}}_{L^{b}} < +\infty.
    \quad
    \blacksquare
\end{equation}

Our next job is to extract the small $\Delta M$ from the supremum part.

\begin{Lemma}
    \label{lem:Taylor}
    For each point $(q,p)$, there exists $M_{\ast}\in [M,\tilde{M}]$
      such that
      \begin{equation}
          | \phi_{\rm in}(q,p;\tilde{M}) - \phi_{\rm in}(q,p;M)|
          = \Delta M \left| \dfracp{\phi_{\rm in}}{M}(q,p;M_{\ast}) \right|.
      \end{equation}
\end{Lemma}

\Proof
We first remember that $M,\tilde{M}\in (0,1)$.
The function $\psi(q,p;M)$ is $C^{1}$ with respect to $M$
for $M\in (0,1)$, and $\varphi_{\rm in}(k)$ is $C^{1}$ in $k\in [0,1)$.
$\phi_{\rm in}(q,p;M)$ is hence $C^{1}$ for $M\in (0,1)$.
Thus Taylor theorem proves the lemma.
$\blacksquare$

\vspace*{1em}
Lemma \ref{lem:Taylor} gives
\begin{equation}
    \Delta I_{21}
    \leq 2 \Delta M \norm{F'\circ h}_{L^{1}(\mu_{1})}
    \sup_{(q,p)\in\mu_{1}} \left| \dfracp{\phi_{\rm in}}{M}(q,p;M_{\ast}(q,p))
    \right| .
\end{equation}
The last job is to show that the supremum is finite.

\begin{Lemma}
    \label{lem:sup-I21}
    \hspace*{2em}
    $\displaystyle{  \sup_{(q,p)\in\mu_{1}}  }
    \left| \dfracp{\phi_{\rm in}}{M}(q,p;M_{\ast}(q,p)) \right|
    <\infty.$
\end{Lemma}

\Proof
The concrete form of $\partial_{M}\phi_{\rm in}$ is
\begin{equation}
    \dfracp{\phi_{\rm in}}{M}(q,p;M_{\ast})
    = \dfracp{\varphi_{\rm in}}{k}(\psi(q,p;M_{\ast})) \dfracp{\psi}{M}(q,p;M_{\ast}).
\end{equation}
The derivatives of $\varphi_{\rm in}$ and $\psi$ are
\begin{equation}
    \begin{split}
        \dfracp{\varphi_{\rm in}}{k}(k)
        & = \dfrac{2}{K(k)^{2}} [ E'(k)K(k) - E(K)K'(k) ] \\
        & = \dfrac{-2}{k} \left[
      \left( \dfrac{E(k)}{K(k)} - 1 \right)^{2}
      + \dfrac{k^{2}}{1-k^{2}} \left(\dfrac{E(k)}{K(k)} \right)^{2}
    \right]
    \end{split}
\end{equation}
and
\begin{equation}
    \dfracp{\psi}{M}(q,p;M_{\ast})
    = \dfrac{-p^{2}}{4M_{\ast}\sqrt{2M_{\ast}}\sqrt{p^{2}/2+M_{\ast}(1-\cos q)}}
    = \dfrac{-p^{2}}{8M_{\ast}^{2}k},
\end{equation}
where $k$ must be evaluated at $\psi(q,p;M_{\ast}(q,p))$.
The derivative $\partial_{M}\phi_{\rm in}$ is hence
\begin{equation}
    \dfracp{\phi_{\rm in}}{M}
    = \dfrac{p^{2}}{4M_{\ast}^{2}} \dfrac{1}{k^{2}}
    \left[
      \left( \dfrac{E(k)}{K(k)} - 1 \right)^{2}
      + \dfrac{k^{2}}{1-k^{2}} \left(\dfrac{E(k)}{K(k)} \right)^{2}
    \right].
\end{equation}
The functions $E(k)$ and $1/K(k)$ are finite in the interval $k\in [0,1]$.
Therefore, remembering $M_{\ast}(q,p)\in [M,\tilde{M}]$ and is positive,
it is enough to show
\begin{itemize}
      \item No divergence at $k=0$,
      \item No appearance of $k=1$.
\end{itemize}

\noindent
\underline{No divergence at $k=0$:}
Around $k=0$, from the Taylor expansions of $K(k)$, \eqref{eq:K-Taylor},
and $E(k)$, \eqref{eq:E-Taylor}, we have
\begin{equation}
    \label{eq:EoverK}
    \dfrac{E(k)}{K(k)} = 1 - \dfrac{k^{2}}{2} + O(k^{4}).
\end{equation}
Thus, we have
\begin{equation}
    \begin{split}
        \dfracp{\phi_{\rm in}}{M}
        & = \dfrac{p^{2}}{4M_{\ast}^{2}} \dfrac{1}{k^{2}}
        \left[
          \left( \dfrac{E(k)}{K(k)} - 1 \right)^{2}
          + \dfrac{k^{2}}{1-k^{2}} \left(\dfrac{E(k)}{K(k)} \right)^{2}
        \right] \\
        & = \dfrac{p^{2}}{4M_{\ast}^{2}} \dfrac{1}{k^{2}}
        \left[ O(k^{4}) + \dfrac{k^{2}}{1-k^{2}}(1+O(k^{2}) ) \right] \\
        & \to \dfrac{p^{2}}{4M_{\ast}^{2}} \quad (k\to 0).
    \end{split}
\end{equation}
Actually, $k\to 0$ implies $(q,p)\to (0,0)$
and hence $\partial_{M}\phi_{\rm in}\to 0$.

\vspace*{1em}
\noindent
\underline{No appearance of $k=1$:}
$k=\psi(q,p;M_{\ast})$ is an increasing function of $p$
for a fixed $q$, thus it is enough to investigate the upper value
of $k$ on the upper boundary of $\mu_{1}$:
\begin{equation}
    p = \sqrt{2(2M_{1}-M+M\cos q)}.
\end{equation}
Substituting this $p$ into $\psi(q,p;M_{\ast})$, we have
\begin{equation}
    \begin{split}
        k^{2}
        & = \dfrac{2M_{1}-M+M_{\ast}+(M-M_{\ast})\cos q}{2M_{\ast}}
        \leq \dfrac{2M_{1}-M+M_{\ast}+M_{\ast}-M}{2M_{\ast}}
        = 1-2\dfrac{\Delta M}{M_{\ast}}
        \leq 1 - 2\dfrac{\Delta M}{\tilde{M}} .
    \end{split}
\end{equation}
Thus, $k$ is bounded by a positive number
which is smaller than $1$:
\begin{equation}
    k < \sqrt{1 - 2\dfrac{\Delta M}{\tilde{M}}} < 1.
    \quad
    \blacksquare
\end{equation}

\subsubsection{Outside the separatrix: $\mu_{3}$}
\label{sec:mu3}

\begin{Proposition}
\label{prop:DeltaI23}
Under the hypotheses of Theorem~\ref{theo:main}, $|\Delta I_{23}|$ is small.
\end{Proposition}

\Proof The strategy of the proof is almost the same as for Proposition
\ref{prop:DeltaI21}, but we replace $\varphi_{\rm in}(k)$, introduced
in \eqref{eq:varphi-in}, by
\begin{equation}
    \label{eq:varphi-out}
    \varphi_{\rm out}(k) = \dfrac{2k^{2}E(1/k)}{K(1/k)}-2k^{2}+1.
\end{equation}
We show the following Lemma,
which corresponds to Lemma \ref{lem:sup-I21}:
\begin{Lemma}
    \label{lem:sup-I23}
    \hspace*{2em}
    $\displaystyle{ \sup_{(q,p)\in\mu_{3}}  }
    \left| \dfracp{\phi_{\rm out}}{M}(q,p;M_{\ast}(q,p)) \right|
    <\infty$.
\end{Lemma}

\Proof
The derivative of $\varphi_{\rm out}$ is
\begin{equation}
    \begin{split}
        \dfracp{\varphi_{\rm out}}{k}
        & = 4k \left( \dfrac{E(1/k)}{K(1/k)} - 1 \right)
        - \dfrac{2}{K(1/k)^{2}} \left[
          E'(1/k)K(1/k) - E(1/k)K'(1/k) \right] \\
        & = 4k \left( \dfrac{E(1/k)}{K(1/k)} - 1 \right)
        - 2k \left[ \left( \dfrac{E(1/k)}{K(1/k)} - 1 \right)^{2}
          + \dfrac{1}{1-k^{2}} \left(\dfrac{E(1/k)}{K(1/k)} \right)^{2} \right].
    \end{split}
\end{equation}
The value of $k$ must be evaluated at $k=\psi(q,p;M_{\ast})$.
The derivative $\partial_{M}\phi_{\rm out}$ is hence
\begin{equation}
    \dfracp{\phi_{\rm out}}{M}
    = \dfrac{p^{2}}{4M_{\ast}^{2}} \left[
      \left( \dfrac{E(1/k)}{K(1/k)} - 1 \right)^{2}
      + \dfrac{1}{1-k^{2}} \left(\dfrac{E(1/k)}{K(1/k)} \right)^{2}
      - 2 \left( \dfrac{E(1/k)}{K(1/k)} - 1 \right)
    \right].
\end{equation}
The functions $E(1/k)$ and $1/K(1/k)$ are finite
in the interval $k\in [1,\infty]$,
and hence it is enough to show
\begin{itemize}
      \item No appearance of $k=1$,
      \item No divergence at $k=\infty$.
\end{itemize}

\noindent
\underline{No appearance of $k=1$:}
As commented previously, $\psi$ is an increasing function of $p$,
and hence a lower bound for $k$
is given by considering the lower boundary of $\mu_{3}$
\begin{equation}
    p = \sqrt{2(2M_{2}-M+M\cos q)}.
\end{equation}
Substituting this $p$ into $\psi(q,p;M_{\ast})$, we have
\begin{equation}
    k^{2} = \dfrac{2M_{2}-M+M_{\ast}+(M-M_{\ast})\cos q}{2M_{\ast}}
    \geq \dfrac{2M_{2}-M+M_{\ast}-|M-M_{\ast}|}{2M_{\ast}} = \dfrac{M+2\Delta M}{M_{\ast}} \geq 1+\dfrac{\Delta M}{\tilde{M}}.
\end{equation}
Thus, we have proved
\begin{equation}
    k > \sqrt{ 1 + \dfrac{\Delta M}{\tilde{M}}}  > 1.
\end{equation}

\vspace*{1em}
\noindent
\underline{No divergence at $k=\infty$:}
The estimation \eqref{eq:EoverK} gives,
in the limit $k\to\infty$,
\begin{equation}
    \dfrac{E(1/k)}{K(1/k)} = 1 - \dfrac{1}{2k^{2}} + O(1/k^{4}),
\end{equation}
and $p^{2}\leq 4\tilde{M}k^{2}$.
Thus, we have
\begin{equation}
    \left| \dfracp{\phi_{\rm out}}{M} \right|
    \leq \dfrac{\tilde{M}k^{2}}{M_{\ast}^{2}}
    \left| O(1/k^{4}) + \dfrac{1+O(1/k^{2})}{1-k^{2}}
      + \dfrac{1}{k^{2}} + O(1/k^{4}) \right|
    \to 0, \quad (k\to\infty).
    \quad \blacksquare
\end{equation}

\section{Numerical Tests}
\label{sec:num}

In this section, we present some numerical simulations of the
Vlasov equation for the HMF model, using a semi-Lagrangian code
\cite{DeBuyl10}. The purpose is twofold:\\
i) Illustrate numerically Theorems 1 and 2. We will show that a
modification of the distribution function small in $W^{s,a}$, with
$1\leq s<1+1/a$ can destabilize a homogeneous stable stationary
state. By contrast, we never observe the destabilization of an
inhomogeneous stationary state by such perturbations.\\
ii) Perform a few numerical tests in a case not covered by Theorem 1,
where modifications are in spaces rougher than $W^{1,a}$.

\subsection{Set up}
\label{sec:setup}

In this section we concentrate on $a=2$,
and denote $W^{s,2}$ by $H^{s}$ following the conventional notation.
Let us consider the following modification of the reference state $f_0$:
\begin{equation}
    g_{\varepsilon,\delta}(q,p)
    = \varepsilon^{\delta} e^{-(h(q,p)-h(0,0))/T\varepsilon^{2}},
    \label{eq:modif}
\end{equation}
where
\begin{equation}
    \label{eq:h-HMF}
    h(q,p) = \dfrac{p^{2}}{2} - M\cos q
\end{equation}
is the one-body Hamiltonian. 
The modification we actually use is slightly different from \eqref{eq:modif},
to ensure the normalization of the modified reference stationary state.
The function $g_{\varepsilon,\delta}(q,p)$ is almost zero
except for a neighborhood of the origin,
thus we may approximate $h(q,p)$ in the inhomogeneous case as
$h(q,p)\simeq (p^{2}+q^{2})/2$.
The exponent $\delta$ controls 
since the $H^{s}$ norm of $g_{\varepsilon,\delta}$: 
\begin{equation}
    \norm{g_{\varepsilon,\delta}}_{H^{s}} \simeq \left\{
      \begin{array}{ll}
          \varepsilon^{1/2+\delta-s}, & \text{homogeneous case,} \\
          \varepsilon^{1+\delta-s}, & \text{inhomogeneous case.}
      \end{array}
    \right.
\end{equation}
Since $H^s\subset H^t$ for $t\leq s$, we have
\begin{equation}
    g_{\varepsilon,\delta} \in \left\{
      \begin{array}{ll}
          H^{s}(s<1/2+\delta), & \text{homogeneous case,} \\
          H^{s}(s<1+\delta),  & \text{inhomogeneous case.} \\
      \end{array}
    \right.
\end{equation}

Let us estimate the contribution of $g_{\varepsilon,\delta}$ to
the stability functional, that is $I[g_{\varepsilon,\delta}]-1$.  The
homogeneous case is straightforward.  For the inhomogeneous case, we
expand $\cos q$ in the Taylor series, and using the angle-action
variables, $q=\sqrt{2J}\sin\Theta$ and $p=\sqrt{2J}\cos\Theta$, we
obtain the following approximation of the function $w(q,p)$:
\begin{equation}
    w(q,p) \simeq (p^{2}+q^{2})^{2} + O[(q,p)^6].
\end{equation}
From the above approximation, we have the estimates
of $I[g_{\varepsilon,\delta}]-1$ both for the homogeneous
and the inhomogeneous cases:
\begin{equation}
    \label{eq:Ig}
    I[g_{\varepsilon,\delta}]-1 = \left\{
      \begin{array}{ll}
          \varepsilon^{\delta-1}, & \text{homogeneous case,} \\
          \varepsilon^{\delta+4},  & \text{inhomogeneous case.} \\
      \end{array}
    \right.
\end{equation}
These estimations imply: i) In homogeneous case, the modification
$g_{\varepsilon,\delta}$ may change the sign of the stability
functional even in the limit $\epsilon\to 0$ for $\delta\leq 1$,
that is when $g_{\varepsilon,\delta}$ is small in $H^{s}(s<1/2+\delta)$.
ii) In inhomogeneous case, no $\delta$ can change the sign of the
stability functional in the limit $\epsilon\to 0$ contrasting with the
homogeneous case.

Based on the above considerations, we prepare a perturbed initial
distribution
\begin{equation}
    f_{\epsilon,\delta,\mu}(q,p) = A \left[ e^{-h(q,p)/T} (1 + \mu\cos q)
      + \varepsilon^{\delta} e^{-(h(q,p)-h(0,0))/T\varepsilon^{2}} \right],
\end{equation}
where $A$ is the normalization factor,
the first term corresponds to the distribution
in thermal equilibrium,
the third term corresponds to
the modification $g_{\varepsilon,\delta}$, and the second term
proportional to $\mu$ is a perturbation to check the stability
of the stationary state $f_{\varepsilon,\delta,0}(q,p)$.  The
magnetization $M$ in the one-body Hamiltonian $h(q,p)$ must satisfies
the self-consistent equation
\begin{equation}
    M = \int f_{\epsilon,\delta,0}(q,p) \cos q dqdp.
\end{equation}
The critical temperature in thermal equilibrium states
is $T_{c}=0.5$ in the HMF model, and therefore,
we will set $T=0.6$ for homogeneous case and $T=0.4$ for inhomogeneous case.

We perform numerical integration of the Vlasov equation
by using the semi-Lagrangian code \cite{DeBuyl10}
with the time step $\Delta t=0.05$.
We introduce a mesh on the truncated phase space
$(q,p)\in ]-\pi,\pi]\times [-3,3]$,
and the mesh size is $512\times 512$ unless otherwise specified.

\subsection{Homogeneous case}
\label{sec:homogeneous-case}

\begin{figure}
    \centering
    \includegraphics[width=8cm]{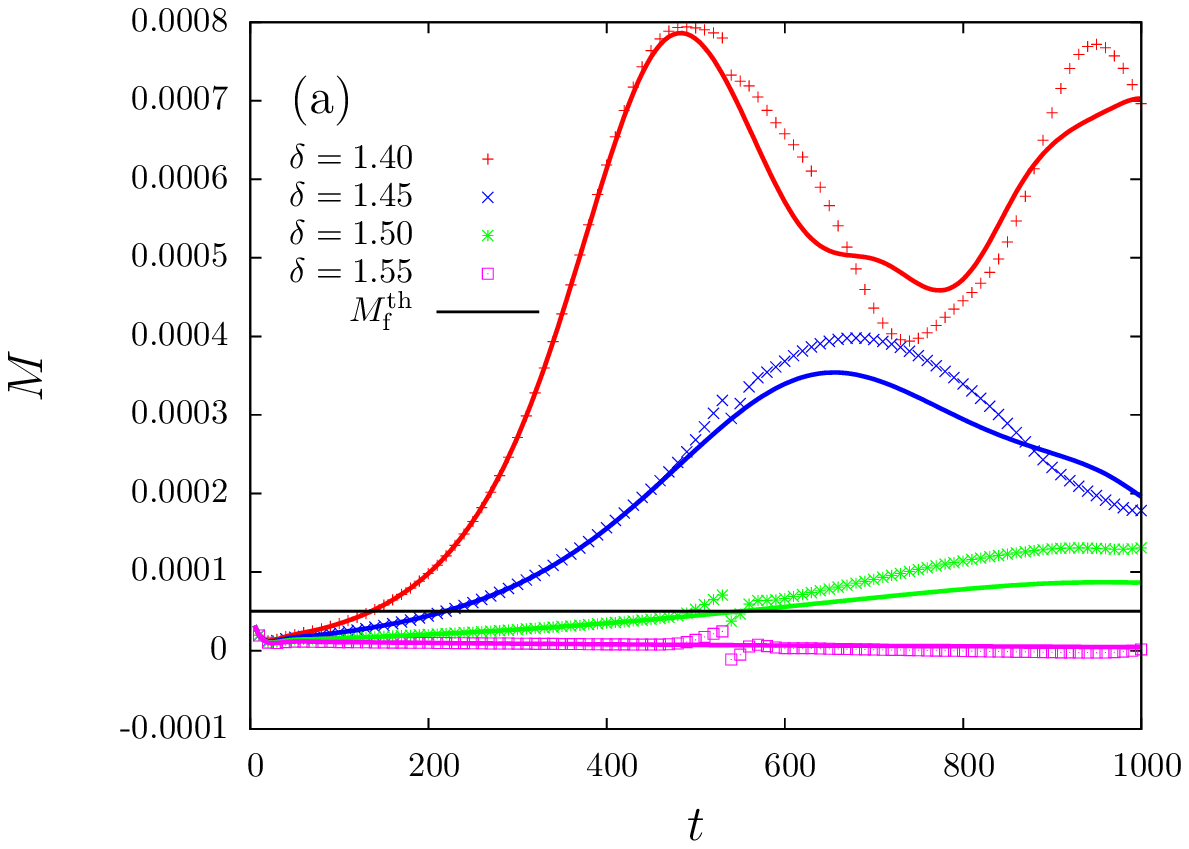}
    \includegraphics[width=8cm]{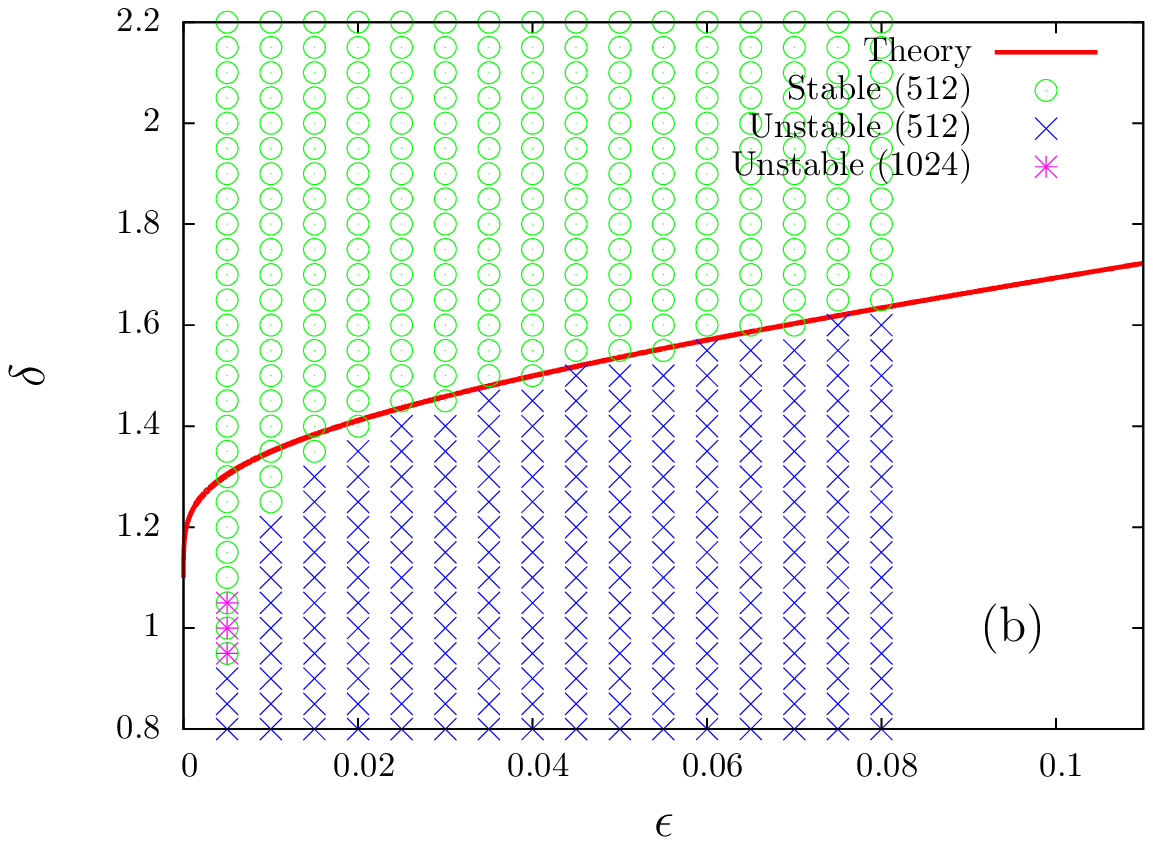}
    \caption{
      (color online)
      (a) Typical temporal evolutions of magnetization $M(t)$.
      $\epsilon=0.05$, which gives the stability threshold value
        as $\delta_{\rm c}\simeq1.536$. The values of $\delta$ are
      $\delta=1.40$ (red plus), $1.45$ (blue crosses), $1.50$ (green stars)
      and $1.55$ (purple boxes),
      with which the solid lines are computed by using
      a finer mesh size $1024\times 1024$ from top to bottom.
      The black straight line marks the level
      $M_{\rm f}^{\rm th}=5.10^{-5}$,
      which is used in the panel (b).
      (b) Phase diagram on the $(\varepsilon,\delta)$ plane.
      Solid red line represent the boundary of stability
      defined by $1-(1+\varepsilon^{\delta-1})/2T(1+\varepsilon^{\delta+1})=0$.
      Green circles and blue crosses represent
      $M_{\rm f}<M_{\rm f}^{\rm th}$ and $M_{\rm f}>M_{\rm f}^{\rm th}$
      respectively.
      Purple stars, overwritten on green circles,
      are for $M_{\rm f}>M_{\rm f}^{\rm th}$,
      but with a finer mesh size $1024\times 1024$.}
    \label{fig:homo}
\end{figure}

We set the magnetization as zero in the one-body Hamiltonian
$h(q,p)$, \eqref{eq:h-HMF}.
Typical temporal evolution of magnetization is exhibited in
Fig.\ref{fig:homo}(a).  In a short time region $M(t)$ decreases, and
then it increases if the considered stationary state is unstable. 
The instability gets weaker as $\delta$ approaches the
threshold value, which can be computed by the stability functional
\begin{equation}
    \label{eq:I-homo}
    I[f_{\varepsilon,\delta,0}]
    = 1 - \dfrac{1+\varepsilon^{\delta-1}}{2T(1+\varepsilon^{\delta+1})}.
\end{equation}
We remark that for the thermal equilibrium with $T=0.6$,
$I[f_{0,\delta,0}]=1/6>0$ and hence the unmodified distribution
$f_{0,\delta,0}$ is stable.  The strange looking discontinuity
around $t=512$ is an artifact due to the mesh size; indeed, it
disappears when a finer $1024\times 1024$ mesh is
used. Nevertheless, in most cases a $512\times 512$ mesh is
sufficient to judge the stability of the modified state.

We use the perturbation level $\mu=10^{-4}$.
Varying $\epsilon$ and $\delta$,
we compute values of magnetization at $t=1000$,
and judge the stability of the modified states $f_{\varepsilon,\delta,0}$.
From the typical temporal evolutions of $M(t)$,
we use the criterion that the state $f_{\varepsilon,\delta,0}$
is unstable if the final magnetization $M_{\rm f}=M(1000)$
is larger than $M_{\rm f}^{\rm th}=\mu/2=5.10^{-5}$,
which is slightly larger than the initial value
$M_{\rm i}= \mu/[2(1+\epsilon^{\delta+1})]$.
The phase diagram on the $(\varepsilon,\delta)$
plane is reported in Fig.\ref{fig:homo}(b), together with the theoretical
threshold line defined by $I[f_{\varepsilon,\delta,0}]=0$.

Numerical results are not in perfect agreement with the
theoretical prediction.  There are three numerical reasons.  1) Mesh
size: A smaller $\varepsilon$ implies that the modification is
strongly concentrated around the line $p=0$.  As a result, we need a
finer mesh to capture the modification for a smaller $\varepsilon$.
Indeed, using the mesh size $1024\times 1024$, 
three points on the line $\varepsilon=0.005$
are found unstable, whereas the $512\times 512$ mesh
judged the same states stable.  2) Computational time: If
$\delta$ goes up to the theoretical line with a fixed $\epsilon$,
the strength of instability gets weaker.  Thus, a longer time computation
is required to observe instability, since typical $M(t)$ curves
decrease in a short time region.  3) Weak instability: in relation with
the point 2), if the instability is very weak, then the magnetization
saturates at a lower level than the threshold $M_{\rm f}^{\rm th}$.

Summarizing,
a large $\delta$, corresponding to a ``smooth'' space $H^{1/2+\delta}$,
keeps the modified state stable,
but a small $\delta$, corresponding to a ``rough'' space
changes the stability and the modified state becomes unstable.
We stress that, even $\varepsilon$ is small enough,
there is a modification which makes the state unstable.

\subsection{Inhomogeneous case}
\label{sec:inhomogeneous-case}

The estimation of the stability functional \eqref{eq:Ig},
suggests that a small modification by $g_{\epsilon,\delta}$
cannot change the stability of the inhomogeneous stationary state
$f_{0,\delta,0}$.
We numerically confirm this suggestion.

Computations are performed along two lines: (i) $\delta=0.5$. (ii)
$\epsilon=0.05$.  We choose the value $\delta=0.5$ since it gives the
same threshold $s=3/2$ as the homogeneous case with $\delta=1$.
Then, we examine stability by decreasing $\delta$, which means
that the modification $g_{\epsilon,\delta}$ becomes ``rough''.

\begin{figure}
    \centering
    \includegraphics[width=8cm]{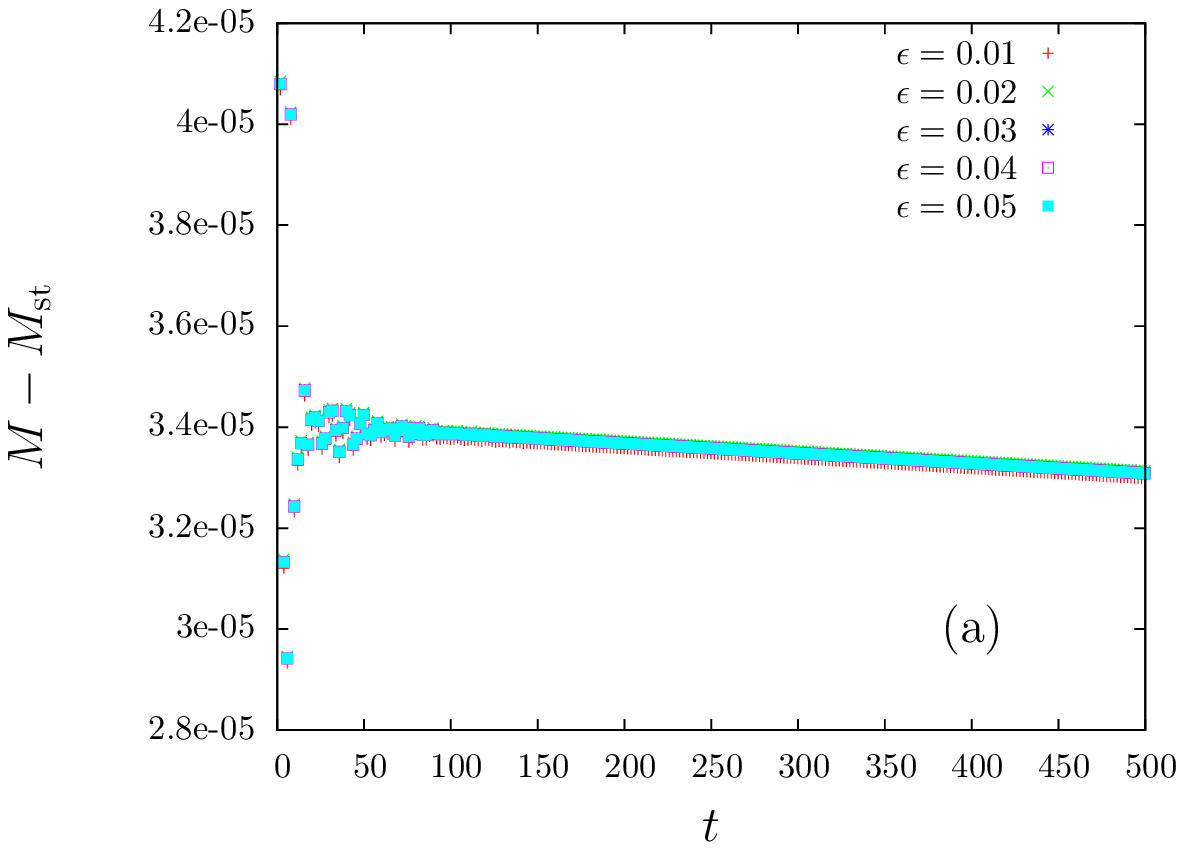}
    \includegraphics[width=8cm]{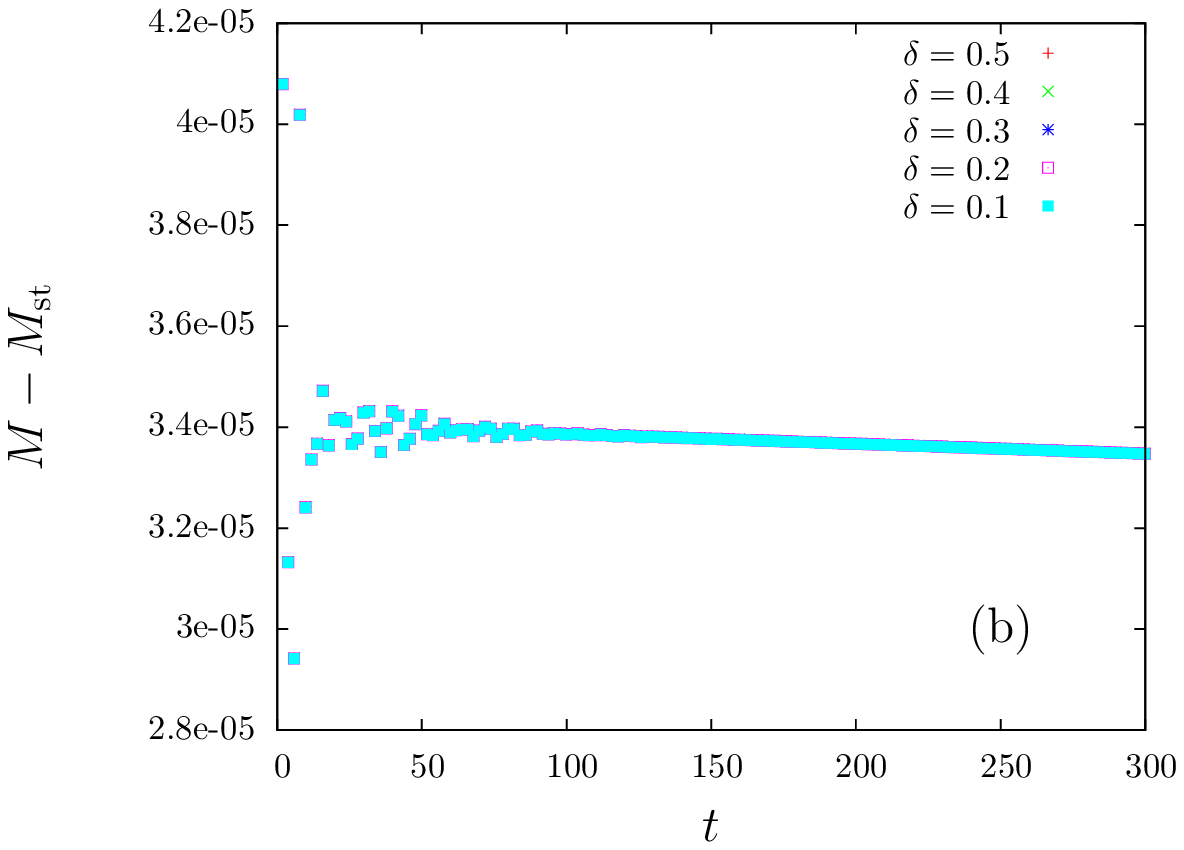}
    \caption{
      (color online)
      Temporal evolutions of magnetization $M(t)-M_{\rm st}$,
      where $M_{\rm st}$ is the value satisfying the self-consistent equation.
      (a) $\delta=0.5$, $\varepsilon=0.01$ to $0.05$.
      (b) $\epsilon=0.05$, $\delta=0.5$ to $0.1$.
      In both panels, $\mu=10^{-4}$
      and five types of points corresponding
      to five values of $\varepsilon$ or $\delta$
      almost collapse.}
    \label{fig:inhomo}
\end{figure}

Along the two lines, no instability is observed.  Thus, the stability
of the inhomogeneous stationary state does not seem to change even
when the modification is ``rough'' enough.  We have used a
perturbation level $\mu=10^{-4}$; changing it do not significantly
affect the results.

\section{Conclusion and Discussions}
\label{sec:concl}

We have shown that, in the 1D Vlasov equation with a cosine potential
(corresponding to the HMF model), any non homogeneous stable
monotonous stationary has a neighborhood in the $W^{1,a}(a>2)$ norm
that does not contain any unstable monotonous stationary states with
nearby magnetization.

This is in striking contrast with the homogeneous case, where all 
neighborhoods of a reference stable state in norms controlling only one 
derivative do contain unstable stationary states.

These results are illustrated with direct simulations of the Vlasov
equation, using a reference stationary state and controlling the norm
of a modification of this reference state in various Sobolev spaces.

Theorem~\ref{theo:main} points to an important difference in the
mathematical structure of the neighborhoods of homogeneous and
inhomogeneous stationary states of Vlasov equation.  Understanding the
physical consequences of this fact, especially with respect to the non
linear evolution of a perturbation, remains an open question.

Finally, we stress that the proof of Theorem~\ref{theo:main} relies on
the knowledge of a stability functional which is rather simple in the
HMF model.  Extending the Theorem to other models having long-range
interactions, where such a simple stability functional is not
available, is another open problem.

\appendix
\section{$W^{1,p}$ spaces}

For $X\subset\mathbb{R}^{n}$, the Sobolev space $W^{1,p}(X)$ is defined by
\begin{equation}
    \label{eq:sobolev}
    W^{1,p} = \left\{ f: X\to \mathbb{R} ~\left|~ \norm{f}_{L^{p}}+\norm{\nabla f}_{L^{p}} < \infty \right. \right\} .
\end{equation}
The norm on $W^{1,p}$ is:
\begin{equation}
    \label{eq:sobolev-norm}
    \norm{f}_{W^{1,p}} = \norm{f}_{L^{p}}+\norm{\nabla f}_{L^{p}}
\end{equation}
where we recall
\begin{equation}
    \label{eq:Lp}
    \norm{f}_{L^{p}} = \left\{
      \begin{array}{ll}
          \displaystyle{
            \left( \int_{X} |f(x)|^{p} dx \right)^{1/p}
          }
          & (1\leq p<\infty) \\
          \displaystyle{
            \sup_{x\in X} |f(x)|
          }
          & (p=\infty).
      \end{array}
      \right.
\end{equation}
We refer the reader to \cite{FracSob} for fractional Sobolev spaces, needed for Theorem \ref{theo:hom}. 

\section{Some useful properties of the complete elliptic integrals}
\label{sec:cei}
We list a few useful properties of the complete elliptic integrals.
\begin{enumerate}
      \item $K$ is monotonically increasing,
    and $E$ is monotonically decreasing
      \item $K(0)=E(0)=\pi/2$
      \item $K(k)\to\infty~(k\to 1)$
      \item $E(1)=1$
      \item The derivatives of $K$ and $E$ are
    \begin{equation}
        \dfracd{K}{k}(k) = \dfrac{E(k)-(1-k^{2})K(k)}{k(1-k^{2})}
    \end{equation}
    and
    \begin{equation}
        \dfracd{E}{k}(k) = \dfrac{E(k)-K(k)}{k}.
    \end{equation}
      \item Taylor expansions of $K$ and $E$ around $k=0$ are
    \begin{equation}
        \label{eq:K-Taylor}
        K(k) = \dfrac{\pi}{2} \left(
          1 + \dfrac{k^{2}}{4} + \dfrac{9}{64}k^{4} + \cdots \right),
    \end{equation}
    and
    \begin{equation}
        \label{eq:E-Taylor}
        E(K) = \dfrac{\pi}{2} \left(
          1 - \dfrac{k^{2}}{4} -\dfrac{9}{192}k^{4} - \cdots \right).
    \end{equation}
\end{enumerate}


\begin{thebibliography}{99}
\bibitem{Landau46} L. Landau,
        {\it J.~Phys.~USSR}~{\bf 10}, 25 (1946).
\bibitem{BGK} I. B. Bernstein, J. M. Greene, and M. D. Kruskal,
        {\it Phys. Rev.}~{\bf 108}, 546-550 (1957).
\bibitem{ONeil65} T. M. O'Neil, 
	{\it Phys. Fluids}~{\bf 8}, 2255 (1965). 
\bibitem{Holloway91}
	J. P. Holloway and J. J. Dorning, 
	{\it Phys. Rev. A}~{\bf 44}, 3856 (1991).
\bibitem{Lancellotti98}	C. Lancellotti and J. J. Dorning, 
	{\it Phys. Rev. Lett.}~{\bf 80}, 5236 (1998).
\bibitem{Manfredi97} G. Manfredi,
	{\it Phys. Rev. Lett.}~{\bf 79}, 2815 (1997).
\bibitem{Mouhot11} C. Mouhot and C. Villani,
        {\it Acta Mathematica}~{\bf 207}, 29 (2011).
\bibitem{Mouhot10} C. Mouhot and C. Villani,
        {\it J.~Math.~Phys.}~{\bf 51}, 015204 (2010).
\bibitem{Lin11} Z. Lin and C. Zeng,
        {\it Comm. Math. Phys.}~{\bf 306}, 291-331 (2011).
\bibitem{BOY11} J. Barr\'e, A. Olivetti A and Y. Y. Yamaguchi,
        {\it J. Phys. A}~{\bf 44}, 405502 (2011).
\bibitem{BY13} J. Barr\'e and Y. Y. Yamaguchi,
        {\it J. Phys. A}~{\bf 46}, 225501 (2013).
\bibitem{Morita11} F. Bouchet and H. Morita,
        {\it Physica D}~{\bf 239}, 948 (2010).
\bibitem{BinneyTremaine} J. Binney and S. Tremaine,
        {\it Galactic Dynamics Second Edition},
        2008 (Princeton University Press).
\bibitem{Ogawa13} S. Ogawa,
        {\it Phys. Rev. E}~{\bf 87}, 062107 (2013).
\bibitem{Lemou12} M. Lemou, F. M\'ehats and P. Rapha\"el,
        {\it Inventiones Mathematicae}~{\bf 187}, 145 (2012).
\bibitem{FracSob}  E. Di Nezza, G. Palatuccia, E. Valdinoci,
         {\it Bulletin des Sciences Math{\'e}matiques}~{\bf 136}, 521(2012). 
\bibitem{DeBuyl10} P. de Buyl,
         {\it Commun. Nonlinear Sci. Numer. Simulat.}~{\bf 15}, 2133 (2010).
\end{thebibliography}
\end{document}